\documentclass[aip,amsmath,amssymb,reprint]{revtex4-2}

\usepackage{graphicx}
\usepackage[T1]{fontenc}
\usepackage[english]{babel}

\makeatletter
\def\@email#1#2{%
 \endgroup
 \patchcmd{\titleblock@produce}
  {\frontmatter@RRAPformat}
  {\frontmatter@RRAPformat{\produce@RRAP{*#1\href{mailto:#2}{#2}}}%
    \frontmatter@RRAPformat}
  {}{}
}%
\makeatother

\DeclareMathOperator{\erf}{erf}
\DeclareMathOperator{\erfi}{erfi}

\DeclareMathOperator{\erfc}{erfc}
\DeclareMathOperator{\arctan_var}{arctan}

\newcommand{\sfrac}[2]{\mbox{\small$\displaystyle\frac{#1}{#2}$}}
\newcommand{\pheq}{\mathrel{\phantom{=}}}
\DeclareMathAlphabet{\vecfont}{OT1}{cmr}{bx}{it}
\renewcommand{\vec}[1]{\vecfont{#1}}
\newcommand{\grad}{\boldsymbol{\nabla}}  

\begin{document}

\title{Contactium: A strongly correlated model system}

\author{Jerzy Cioslowski}
\thanks{To whom all the correspondence should be addressed, %
        e-mail: jerzy@wmf.univ.szczecin.pl}
\affiliation{Institute of Physics, University of Szczecin, Wielkopolska 15, %
  \mbox{70-451 Szczecin, Poland}}

\author{Berthold-Georg~Englert}
\affiliation{Centre for Quantum Technologies, %
  National University of Singapore, 3 Science Drive 2, %
  Singapore 117543, Singapore}
\affiliation{Department of Physics, %
  National University of Singapore, 2 Science Drive 3, %
  Singapore 117542, Singapore}
\affiliation{MajuLab, CNRS-UCA-SU-NUS-NTU %
  International Joint Research Unit, Singapore}
\author{Martin-Isbj\"orn~Trappe}
\affiliation{Centre for Quantum Technologies, %
  National University of Singapore, 3 Science Drive 2, %
  Singapore 117543, Singapore}
\author{Jun~Hao~Hue}
\affiliation{Centre for Quantum Technologies, %
  National University of Singapore, 3 Science Drive 2, %
  Singapore 117543, Singapore}

\date[]{Posted on the arXiv on 27 March 2023}

\begin{abstract}At the limit of an infinite confinement strength $\omega$, the
  ground state of a system that comprises two fermions or bosons in a harmonic
  confinement interacting through the Fermi--Huang pseudopotential remains
  strongly correlated.
  A detailed analysis of the one-particle description of this ``contactium''
  reveals several peculiarities that are not encountered in conventional model
  systems (such as the two-electron harmonium atom, ballium, and spherium)
  involving Coulombic interparticle interactions.
  First of all, none of the natural orbitals (NOs)
  $\{ \psi_\mathfrak{n}(\omega;\vec r) \}$ of the contactium is unoccupied,
  which implies nonzero collective occupancies for all the angular momenta.
  Second, the NOs  and their nonascendingly ordered occupation numbers $\{
  \nu_\mathfrak{n} \}$ turn out to be related to the eigenfunctions and
  eigenvalues of a zero-energy Schr\"odinger equation with an attractive
  Gaussian potential.
  This observation enables the derivation of their properties such as the
  $\mathfrak{n}^{-4/3}$ asymptotic decay of $\nu_\mathfrak{n}$ at the
  $\mathfrak{n} \to \infty$ limit (which differs from that of
  $\mathfrak{n}^{-8/3}$ in the Coulombic systems), the independence of the
  confinement energy
  ${v_\mathfrak{n} = \langle  \psi_\mathfrak{n}(\omega;\vec r) | \frac{1}{2} %
  \omega^2r^2 | \psi_\mathfrak{n}(\omega;\vec r) \rangle}$ of
  $\mathfrak{n}$, and the $\mathfrak{n}^{-2/3}$ asymptotic decay of the
  respective contribution $\nu_\mathfrak{n}t_\mathfrak{n}$ to the kinetic
  energy.
  Upon suitable scaling, the weakly occupied NOs of the contactium turn out to
  be virtually identical with those of the two-electron harmonium atom at the
  ${\omega \to \infty}$ limit, despite the entirely different interparticle
  interactions in these systems.  
\end{abstract}

\maketitle

\section{Introduction}
Interactions between particles confined by external potentials introduce
correlations that limit the accuracy of simple descriptions (such as the
Hartree--Fock and Gross--Pitaevskii approximations for fermions and bosons,
respectively) of their quantum states.
This limitation is lift\-ed upon employment of more sophisticated formalisms
based upon quantities such as the one-particle density matrix (the 1-matrix),
the one-particle density, and var\-i\-ous one-particle functions (commonly called
orbitals).  In fact, in the case of Coulombic systems (i.e. atoms, ions, and
molecules), formalisms that involve these quantities are the mainstay of
approaches to the electron correlation problem.\cite{1}
Development and implementation of these approaches is greatly aided by
benchmarking on model systems for which exact wave functions can be written in
closed forms. 

One of such systems is the harmonium atom (also known as hookium or Hooke's
atom) that comprises Cou\-lombically interacting fermions in a harmonic external
potential.
Its two-particle version,\cite{2,3,4,5} introduced over six\-ty years
ago,\cite{6} has been extensively used (especially with the confinement
strengths that correspond to compact wave functions) in conjunction with
diverse formalisms of quantum chemistry.\cite{7,8}
Other model systems, such as spherium (i.e. two electrons restricted to the
surface of a sphere\cite{9}) and ballium (i.e. two electrons trapped in a
spherical box\cite{10}) have also been considered but found only limited
applications due to the fact that, unlike harmon\-ium (which describes
three-dimensional quantum dots), they do not pertain to any experimentally
\mbox{realizable} phys\-i\-cal system. 

The 1-matrix functional theory (1-RDMFT) is an emergent approach to the
accurate treatment of correlations in species composed of either
fermions\cite{11} or bosons\cite{12} that holds a promise of becoming an
alternative to density-functional theory (DFT).
When formulated in terms of natural orbitals (NOs) 
$\{ \phi_i(\vec r) \}$, \mbox{1-RDMFT} leads to (in principle exact, in
practice approximate) functionals for the correlated component of the
interparticle interaction energy that are ``superuniversal,'' i.e.\ it can be
expressed solely in terms of the occupation numbers of the NOs, some
spin-related quantities, and the two-particle matrix elements of the
interaction potential, such as the two-electron integrals  
$\{\langle ij|kl \rangle \} \equiv \bigl\{ \bigl\langle \phi_i(\vec r_1) 
\phi_j(\vec r_2) \big| |\vec r_1-\vec r_2|^{-1} \big|\phi_k(\vec r_1) 
\phi_l(\vec r_2) \bigr\rangle \bigr\}$.\cite{13}
In light of this property, one is tempted to apply this formalism to systems
involving the so-called contact (or zero-range) interactions by simply
replacing $|\vec r_1-\vec r_2|^{-1}$ with $\delta(\vec r_1-\vec r_2)$, where
$\delta(\vec r)$ is the three-dimensional Dirac delta function.
Unfortunately, this simple replacement is not valid due to complications
inherent in the underlying Hamiltonian. 

The lack of the self-adjointness of Hamiltonians with potential energy terms
proportional to $\delta(\vec r)$ was first recognized by Fermi\cite{14} and
then further elaborated by Huang and Yang\cite{15} who concluded that a
regularization of $\delta(\vec r)$ is needed.
The resulting Fermi--Huang pseudo\-potential is now widely used in the
description of ultracold atomic and molecular systems in traps.\cite{16} 

The expressions for the ground-state and excited-state wave functions of a system
of two particles (either fermi\-ons or bosons) whose potential energy comprises
contributions from a harmonic confinement and a term proportional to the
Fermi--Huang pseudopotential are known in closed forms.\cite{17}
Interestingly, the kinetic and interparticle interaction energy components
pertaining to these wave functions are infinite and thus cannot be considered
separately.
This observation prompts the question whether standard approaches (such as DFT
and \mbox{1-RDMFT}) based upon one-particle quantities are applicable to this new
type of a model system for which, in analogy to its Coulombic counterparts
(i.e.\ harmonium, spherium, and ballium), the name ``contactium'' is coined
here.  

In this paper, the one-particle description of the contactium is investigated
and its peculiarities are un\-cov\-ered.
Although two parameters (i.e. the confinement strength $\omega$ and the
coupling constant $\beta$ that multiplies the regularized Dirac delta in the
Fermi-Huang pseudopotential) control the properties of this system, a simple
scaling eliminates one of them.
In order to fa\-cil\-i\-tate a direct comparison with the properties of the
two-electron harmonium atom, $\omega$ is varied while $\beta$ is set to one
(which corresponds to a positive-valued scattering length\cite{17}).
The main focus of the present study is the ${\omega \to \infty}$ limit at which
the contactium remains strongly correlated.

\section{Theory}\label{sec:th}
The  normalized spatial component
\begin{align}\label{a1}
  \Psi(\omega;\vec r_1,\vec r_2)
  &= \frac{\omega^{5/4} \, \Gamma (-\nu)} {2^{3/4} \, \pi^{9/4} \, 
\bigl[ \psi(-\nu)-\psi(-\nu-\frac{1}{2}) \bigr]^{1/2}} \nonumber \\
&\pheq \times \exp\Bigl(-\sfrac{1}{2} \, \omega \, (r_1^2+r_2^2)\Bigr) \nonumber\\ 
  &\pheq\times\frac{U \Bigl( -\nu-\sfrac{1}{2},\sfrac{1}{2},
    \sfrac{1}{2} \, \omega \, r_{12}^2 \Bigr)}{r_{12}} 
\end{align}
of the ground-state wave function of the system under study is an
eigenfunction of the nonrelativistic Hamiltonian 
\begin{equation}\label{a2}
  H = - \frac{1}{2}\bigl(\grad^2_1+\grad^2_2\bigr)
  + \frac{1}{2}\omega^2\bigl(r_1^2+r_2^2\bigr) 
 + \delta_{\rm{reg}}(\vec r_{12})  \quad  
\end{equation}
that involves the Fermi--Huang pseudopotential 
$\delta_{\rm{reg}}(\vec r) = \delta(\vec r) \, \frac{\vec r}{|\vec r|} \cdot
\grad \, |\vec r|$.\cite{14,15,17}
Here and in the following,  $\omega$ is the confinement strength,
${\vec r_{12}=\vec r_1-\vec r_2}$, ${r_1 = |\vec r_1|}$,
${r_2 = |\vec r_2|}$, ${r_{12} = |\vec r_1-\vec r_2|}$,
$\Gamma(t)$ and $\psi(t)= [\ln \Gamma(t)]'$ are the gamma and digamma
functions, respectively,  $U(a,b,t)$ is the \mbox{Tricomi} confluent hypergeometric
function, and $\nu \equiv \nu(\omega)$ is the negative-valued solution of the
equation 
\begin{equation}\label{a3}
  \frac{\Gamma(-\nu-\frac{1}{2})}{\Gamma(-\nu)}
  =  \frac{\omega^{1/2}}{2^{3/2} \, \pi}  \,.
\end{equation}
The energy $E(\omega)$ corresponding to $\Psi(\omega;\vec r_1,\vec r_2)$
equals $(3+2 \nu) \, \omega$.

Since
\begin{align}\label{a4}
  &\pheq
  U \Bigl(-\nu-\sfrac{1}{2},\sfrac{1}{2},\sfrac{1}{2}\omega r_{12}^2 \Bigr) 
    \nonumber\\
  &= \frac{\pi^{1/2}}{\Gamma(-\nu)}
    - \frac{(2 \, \pi)^{1/2}}{\Gamma(-\nu-\frac{1}{2})}
    \omega^{1/2} r_{12} +\cdots
\end{align}
as ${r_{12} \to 0}$, Eq.~(\ref{a1}) reveals the leading singularity in
$\Psi(\omega;\vec r_1,\vec r_2)$ at ${\vec r_1 \to\vec r_2}$.
This singularity persists as ${\omega \to \infty}$, where 
${\nu \to -\frac{1}{2}}$, ${E(\omega) \to 2\omega}$, and 
\begin{align}\label{a5}
  \Psi(\omega;\vec r_1,\vec r_2)
  &\to \Psi_\infty(\omega;\vec r_1,\vec r_2)\nonumber\\
  &=\frac{\omega} {\pi^{3/2}}
    \exp\Bigl(-\sfrac{1}{2} \omega\bigl(r_1^2+r_2^2\bigr)\Bigr)
    \Bigl( \frac{1}{r_{12}}-4\pi \Bigr) \,.
\end{align}
The leading term $\Psi_\infty(\omega;\vec r_1,\vec r_2)$ of the asymptotics
(\ref{a5}) is employed in the following considerations.

\subsection{The natural orbitals and their occupation numbers}\label{sec:th-A}
Since $\Psi_\infty(\omega;\vec r_1,\vec r_2)$ is totally symmetric, the
square-normalized natural orbitals (NOs) $\{\psi_{nlm}(\omega;\vec r)\}$,
which are eigenfunctions of the homogeneous Fredholm equation of the second
kind\cite{18} 
\begin{equation}\label{a6}
  \int \Psi_\infty(\omega;\vec r_1,\vec r_2)\psi_{nlm}(\omega;\vec r_2)
  \, d^3\vec r_2 = \lambda_{nl}\psi_{nlm}(\omega;\vec r_1) \,,
\end{equation}
are given by products of real-valued, square-normalized radial components
$\{\phi_{nl}(\omega;r)\}$ and angular factors that are real-valued combinations
of the respective spherical harmonics $Y_l^{-m}(\theta,\varphi)$ and
$Y_l^m(\theta,\varphi)$.
The cor\-re\-spond\-ing $m$-independent eigenvalues $\{\lambda_{nl} \}$, which
are in\-dexed in a nonascending order by $n=1,2,\dots$, are the real-valued
natural amplitudes (NAs).
Since ${\grad^2_1\frac{1}{|\vec r_1-\vec r_2|}}=
- 4 \pi{\delta(\vec r_1-\vec r_2)}$,
combining Eqs.~(\ref{a5}) and (\ref{a6}) produces
\begin{align}\label{a7}
  &\pheq\biggl(-\frac{1}{2}\grad^2
  -\frac{2\omega} {\pi^{1/2} \lambda_{nl}} \exp\bigl(-\omega r^2\bigr)
  \biggr)\nonumber\\
  &\pheq\times    \Bigl[\exp \Bigl(\sfrac{1}{2}\omega r^2 \Bigr) 
    \psi_{nlm}(\omega;\vec r) \Bigr]=0\,,
\end{align}
a zero-energy Schr\"odinger equation in which
$\exp\big(\frac{1}{2}\omega r^2\big)$ $\psi_{nlm}(\omega;\vec r)$ and  
$-\pi^{-1/2}(2\omega/\lambda_{nl})\exp\bigl(-\omega r^2\bigr)$
play the roles of the wave function and the attractive potential,
respectively.

The number $N_l(\kappa)$ of the $l$-wave bound states of the spherically
symmetric Hamiltonian  
${-\frac{1}{2}\grad^2 - \kappa  V(r)}$, \mbox{where} ${V(r) \ge 0}$ for all $r$,
is known to conform to the asymptotic identity\cite{19} 
\begin{equation}\label{a8}
  \lim_{\kappa \to \infty} \kappa^{-1/2} N_l(\kappa)
  = \frac{2^{1/2}}{\pi} \int_0^{\infty} V(r)^{1/2} \, dr \,.
\end{equation} 
Application of this identity to the present case leads to the conclusion that,
for a given $l$ and $m$, the large-$n$ asymptotic estimates $\{
\tilde\lambda_{nl} \}$ of the NAs read 
\begin{equation}\label{a9}
\tilde\lambda_{nl} = \frac{2}{\pi^{3/2}}\, n^{-2}\,.
\end{equation} 
Similarly, the identity\cite{20}
\begin{equation}\label{a10}
    \lim_{\kappa \to \infty}   \kappa^{-3/2}  N(\kappa)
    = \frac{2^{1/2}}{3 \pi^2} \, \int V(\vec r)^{3/2} \, d^3\vec r   
\end{equation}
for the number $N(\kappa)$ of bound states of the Hamiltonian
${- \frac{1}{2} \grad^2 - \kappa V(\vec r)}$, where ${V(\vec r) \ge 0}$ for
all $\vec r$, produces the power law  
\begin{eqnarray}\label{a11}
  \widetilde\lambda_{\mathfrak{n}}
  = \frac{2^{7/3}}{3^{5/3} \pi^{5/6}} \, \mathfrak{n}^{-2/3}\quad
\end{eqnarray}
for the large-$\mathfrak{n}$ asymptotic estimates
$\{\widetilde\lambda_{\mathfrak{n}}\}$ of the NAs
$\{\lambda_{\mathfrak{n}}\}$ [pertaining to the NOs  
$\{ \psi_{\mathfrak{n}}(\omega;\vec r_1) \}$] indexed in a nonascending order
by ${\mathfrak{n}=1, 2,\dots}$  regardless of $l$ and $m$.
The analogous estimates $\{ \widetilde\nu_{nl} \}$ and 
$\{\widetilde\nu_{\mathfrak{n}} \}$ of the $m$-independent occupation numbers
${\{\nu_{nl}\} \equiv \{\lambda_{nl}^2\}}$ and 
${\{\nu_{\mathfrak{n}}\}\equiv \{\lambda_{\mathfrak{n}}^2 \}}$ follow trivially. 

\begin{widetext}
There are several equivalent expressions for the NAs.  First of all,
Eq.~(\ref{a6}) readily yields
${\lambda_{nl} = \langle \psi_{nlm}(\omega;\vec r_1)|}$
${\Psi_\infty(\omega;\vec r_1,\vec r_2)| \psi_{nlm}(\omega;\vec r_2)\rangle}$.
Second, Eq.~(\ref{a7}) implies
\begin{equation}\label{a12}
\lambda_{nl} = \frac{2  \omega} {\pi^{1/2}} 
\frac{\langle \psi_{nlm}(\omega;\vec r)
  |\exp\bigl(-\omega \, r^2\bigr) |
  \psi_{nlm}(\omega;\vec r) \rangle}
{\bigl\langle \psi_{nlm}(\omega;\vec r) \big|
  \exp \bigl(-\frac{1}{2}\omega r^2 \bigr)  T  
  \exp \bigl(\frac{1}{2}\omega r^2 \bigr) \big|
  \psi_{nlm}(\omega;\vec r)  \bigr\rangle}
=  \frac{2\omega} {\pi^{1/2}} \frac{u_{nl}}{t_{nl}-v_{nl}} \,,
\end{equation}
where $T$ is the kinetic energy operator and the $m$-independent expectation
values read, respectively, 
${u_{nl} = \langle \psi_{nlm}(\omega;\vec r) |}$
${\exp\bigl(-\omega r^2\bigr) | \psi_{nlm}(\omega;\vec r) \rangle}$,
${t_{nl} = \langle \psi_{nlm}(\omega;\vec r)|T|
  \psi_{nlm}(\omega;\vec r)\rangle}$, and 
${v_{nl}=\langle\psi_{nlm}(\omega;\vec r)|\frac{1}{2}\omega^2 r^2|
  \psi_{nlm}(\omega;\vec r)\rangle}$.
Third, since 
\begin{align}\label{a13}
  2\omega\lambda_{nl}
  &= 2\omega\bigl\langle\psi_{nlm}(\omega;\vec r_1)\psi_{nlm}(\omega;\vec r_2) 
    \big| \sum_{\mathfrak{n}=1}^{\infty} \lambda_{\mathfrak{n}}
    \psi_{\mathfrak{n}}(\omega;\vec r_1) \psi_{\mathfrak{n}}(\omega;\vec r_2) 
    \bigr\rangle
    \nonumber \\
  &= 2 \omega\langle \psi_{nlm}(\omega;\vec r_1)\psi_{nlm}(\omega;\vec r_2)|
    \Psi_\infty(\omega;\vec r_1,\vec r_2) \rangle
  = \langle \psi_{nlm}(\omega;\vec r_1)\psi_{nlm}(\omega;\vec r_2)|  H
    |\Psi_\infty(\omega;\vec r_1,\vec r_2)\rangle \nonumber \\
  &= \bigl\langle \psi_{nlm}(\omega;\vec r_1)\psi_{nlm}(\omega;\vec r_2)
    \big|H\big| \sum_{\mathfrak{n}=1}^{\infty}
    \lambda_{\mathfrak{n}}\psi_{\mathfrak{n}}(\omega;\vec r_1)
    \psi_{\mathfrak{n}}(\omega;\vec r_2) \bigr\rangle \nonumber \\
  &= 2\lambda_{nl} (t_{nl}+v_{nl})+ \langle \psi_{nlm}(\omega;\vec r_1)
    \psi_{nlm}(\omega;\vec r_2)|\delta_{\rm{reg}}(\vec r_{12})
   |\Psi_\infty(\omega;\vec r_1,\vec r_2) \rangle  \nonumber \\
  &= 2\lambda_{nl}(t_{nl}+v_{nl}) - \frac{4\omega} {\pi^{1/2}} u_{nl}\,, 
\end{align}\end{widetext}
one has
\begin{eqnarray}\label{a14}
  \lambda_{nl} =\frac{2\omega} {\pi^{1/2}}\frac{u_{nl}}
  {t_{nl}+v_{nl}-\omega} \,. 
\end{eqnarray}
The identities (\ref{a12}) and (\ref{a14}) can be reconciled only if
${v_{nl} =\frac{1}{2}\omega}$ for all $n$ and $l$.
Thanks to this $(nl)$-independence of $v_{nl}$, one has
\begin{align}\label{a15}
  &\pheq\Bigl\langle \Psi_\infty(\omega;\vec r_1,\vec r_2)\Big|
    \sfrac{1}{2}\omega^2 (r_1^2+r_2^2)\Big|
    \Psi_\infty(\omega;\vec r_1,\vec r_2) \Bigr\rangle\nonumber\\ 
  &= 2\sum_{\mathfrak{n}=1}^{\infty} \nu_{\mathfrak{n}} v_{\mathfrak{n}}
    = \omega\sum_{\mathfrak{n}=1}^{\infty} \nu_{\mathfrak{n}} = \omega\,,
\end{align}
as expected.

The large-$n$ asymptotic estimates of $t_{nl}$, $v_{nl}$, and $u_{nl}$ are
available from the general formalism pre\-vi\-ous\-ly applied to systems with
Coulombic two-particle interactions.\cite{21}
They read
\begin{align}\label{a16}
  \widetilde t_{nl} &= \frac{\pi^2}{2}\frac{I_3}{I_1^3}n^2
  = \frac{\pi}{3^{1/2}} \; \omega \, n^2    \,,\\
\label{a17}
\widetilde v_{nl}&= \frac{1}{I_1}\int_0^{\infty}
               \Bigl(\sfrac{1}{2}\omega^2 r^2 \Bigr)
               \exp\Bigl(-\sfrac{1}{2}\omega r^2 \Bigr)\, dr 
              = \frac{1}{2}\omega \,,
\intertext{and}
\label{a18}
\widetilde u_{nl} &= \frac{I_3}{I_1} = 3^{-1/2}   \quad   ,
\end{align}
respectively, where
${I_{\gamma} = \int\limits_0^{\infty}\! \exp\bigl(- 4\omega r^2 )^{\gamma/8} \, dr
= \bigl(\frac{\pi}{2\gamma\omega}\bigr)^{1/2}}$.
The estimate of $v_{nl}$ turns out to be exact, whereas those for $t_{nl}$ and
$u_{nl}$ are consistent with the identities (\ref{a9}), (\ref{a12}), and
({\ref{a14}).
The same formalism yields\begin{widetext}
\begin{equation}\label{a19}
  \widetilde \psi_{nlm}(\omega;\vec r)
  = \Bigl(\sfrac{8\omega}{\pi} \Bigr)^{1/4} J_{l+\frac{3}{2}}(\chi_{nl})^{-1} 
    \frac{\exp\bigl(-\frac{1}{4}\omega r^2\bigr)}{r}
    \erf \big( (\omega/2)^{1/2} r \bigr)^{1/2} 
  J_{l+\frac{1}{2}}
    \Bigl(\chi_{nl}\erf\bigl(\omega/2)^{1/2}r\bigr) \Bigr)
    Y_{lm}(\theta,\varphi) 
\end{equation}
for the large-$n$ asymptotic estimates
$\{ \widetilde \psi_{nlm}(\omega;\vec r) \}$ of the NOs.
In Eq.~(\ref{a19}), $J_l(t)$ is the $l$th Bessel function of the first kind
and $\chi_{nl}$ is the $n$th zero of $J_{l+\frac{1}{2}}(t)$.

\subsection{The 1-matrix and the collective occupancies}\label{sec:th-B}
Let $\Gamma(\omega;\vec r_{1'},\vec r_1)$ be the 1-matrix (per spin)
corresponding to $\Psi(\omega;\vec r_1,\vec r_2)$.
As $\omega \to \infty$,
\begin{equation}\label{a20}
  \Gamma(\omega;\vec r_{1'},\vec r_1)
  \to \Gamma_{\infty}(\omega;\vec r_{1'},\vec r_1) 
  =\frac{\omega^2} {\pi^3}
    \exp\Bigl(-\sfrac{1}{2}\omega\bigl(r_1^2+r_{1'}^2\bigr)\Bigr) 
  \int \frac{\exp\bigl(-\omega r_2^2\bigr)}
    {|\vec r_1-\vec r_2| \, |\vec r_{1'}-\vec r_2|} \, d^3 \vec r_2  \,,
\end{equation}
where $r_{1'}=|\vec r_{1'}|$;
note the absence of the contributions arising from the $-4\pi$ constant term
in the definition of  $\Psi_\infty(\omega;\vec r_1,\vec r_2)$.
Upon insertion of the identity 
${\frac{1}{|\vec r_1-\vec r_2| \, |\vec r_{1'}-\vec r_2|} = 
\frac{1}{\pi} \int_{-\infty}^{\infty} \int_{-\infty}^{\infty}
\exp[-\xi^2(\vec r_1-\vec r_2)^2
-{\xi'}^2(\vec r_{1'}-\vec r_2)^2] \, d\xi \, d\xi'}$,
followed by integration over $\vec r_2$, the change of variables 
$\xi= \zeta^{-1}\bigl[\omega \bigl(1-\zeta^2\bigr)\bigr]^{1/2}\cos\phi$ and 
$\xi'= \zeta^{-1}\bigl[\omega\bigl(1-\zeta^2\bigr)\bigr]^{1/2}\sin\phi$,
and then integration over $\zeta$, Eq.~(\ref{a20}) becomes
\begin{align}\label{a21}
  \Gamma_{\infty}(\omega;\vec r_{1'},\vec r_1)
  &=\frac{2 \,\omega} {\pi^2} \, \int _0^{\pi}
    F(R,r,\vartheta;\phi)^{-1} 
    \exp\Bigl(-\sfrac{1}{2}\omega \bigl(F(R,r,\vartheta;\phi)^2
    -2rR\cos\vartheta \, \cos\phi \bigr) \Bigr) \nonumber \\
  &\pheq\times \biggl( \Re \, \Bigl[\exp\Bigl(-\sfrac{1}{2}i\omega
    F(R,r,\vartheta;\phi)r\sin \vartheta \Bigr) \, 
    \erfi\Bigl( \sfrac{1}{2}\omega^{1/2}
    \bigl(F(R,r,\vartheta;\phi) + ir  \sin\phi\bigr)\Bigr) \Bigr] \nonumber \\
  &\pheq\hphantom{\times\biggl(}
    - \sin\Bigl( \sfrac{1}{2}\omega F(R,r,\vartheta;\phi)r
    \sin\phi\Bigr)\biggr)
    \, d\phi\,,
\end{align}
where ${i =\sqrt{-1}}$, $\Re \, z$ is the real part of $z$,
${R=\frac{1}{2}|\vec r_1+ \vec r_{1'}|}$,
${r=|\vec r_1- \vec r_{1'}|}$, 
${\cos\vartheta =\bigl(r_1^2-r_{1'}^2\bigr)/(2Rr)}$,
$F(R,r,\vartheta;\phi)=\bigl(4R^2+4rR\cos\vartheta\,\cos\phi
+r^2\cos^2\phi\bigr)^{1/2}$, and 
$\erfi(z)$ is the ``imaginary error function'' defined as
${\erfi(z) =- i\erf(iz)}$.
Although the integral that enters Eq.~(\ref{a21}) cannot be evaluated in a
closed form, it is suitable for both numerical calculations and analysis of
the properties of $\Gamma_{\infty}(\omega;\vec r_{1'},\vec r_1)$.
In particular, it readily yields the small-$r$ expansion
\begin{equation}\label{a22}
  \Gamma_{\infty}(\omega;\vec r_{1'},\vec r_1)
  =  \frac{1}{\pi}\omega\exp\bigl(-2\omega R^2\bigr)
  \frac{\erfi\bigl(\omega^{1/2}R\bigr)}{R} 
- \frac{2}{\pi^2} \omega^2 \exp\bigl(-2\omega R^2\bigr) \, r +\cdots
\end{equation}
that, as expected, features a term linear in $r$ (i.e.\ the particle
coalescence cusp).
The presence of this cusp leads to asymptotic decays of
the $m$-independent occupation numbers that are consistent with those given by
Eqs.~(\ref{a9}) and (\ref{a11}).
\end{widetext}

For each azimuthal quantum number $l$, the collective occupancy (per spin and
$m$)  ${\eta_l = \sum_{n=1}^\infty \nu_{nl}}$ is pro\-por\-tion\-al to the norm of
the $l$-wave in the partial-wave decomposition of
$\Psi_\infty(\omega;\vec r_1,\vec r_2)$.\cite{22,23}
These occupancies, which satisfy the sum rule  
${\sum_{l=0}^\infty(2l+1)\eta_l=1}$, are given by the expression
\begin{equation}\label{a23}
  \eta_l= \frac{2(2l+1)\Bigl[\psi\Bigl(\sfrac{2l+3}{4}\Bigr)
    -\psi\Bigl(\sfrac{2l+5}{4}\Bigr) \Bigr]+4}{(2l+1)^2\pi}\,,
\end{equation}
from which the large-$l$ behavior
\begin{equation}\label{a24}
\lim_{l \to \infty} \Bigl(l+\sfrac{1}{2}\Bigr)^3 \eta_l = \frac{1}{2\pi} 
\end{equation} 
is readily deduced.
The collective occupancy 
${\eta_0=2\!-\!4/\pi}$ $\approx 0.726 \, 760$ of the $s$-type NOs
in\-di\-cates the persistence of strong correlation at ${\omega \to \infty}$. 
This limiting value is consistent with the general expression
\begin{eqnarray}\label{a25}
  \eta_0(\omega) = \frac{1}{1+\nu}
  +\frac{\Bigl( \sfrac{\omega}{8\pi^3}\Bigr)^{1/2}-1}{(1+\nu)^2 
\Bigl[ \psi\Bigl(-\nu-\sfrac{1}{2}\Bigr)-\psi(-\nu) \Bigr]} \quad 
\end{eqnarray}
valid for arbitrary $\omega$.
Upon weakening of the confinement, $\eta_0(\omega)$ decreases rapidly, as
revealed by its val\-ues of ${\pi^2/(12\ln 2) - \ln 2 \approx 0.493 \, 422}$,
$(2\pi\ln2 -4)/\bigl(\pi(1-\ln 2)\bigr) \approx 0.368 \, 433$, and 
$(4\ln 2-2)^{-1} - 1 \approx 0.294 \, 350$ computed with Eq.~(\ref{a24})
at $\omega$ equal to $8\pi^3 \approx 248.050$,
$32\pi \approx 100.531$, and $2\pi^3 \approx 62.012\,6$, respectively.

\subsection{Comparison with the strong-confinement limit of the two-electron
  harmonium atom}\label{sec:th-C}
The normalized spatial component of the ground-state wave function of the
two-electron harmonium atom described by the nonrelativistic
Hamiltonian\cite{2,3,4,5,6} 
\begin{equation}\label{a26}
  H = - \frac{1}{2}\bigl(\grad^2_1+\grad^2_2\bigr)
  + \frac{1}{2}\omega^2 \bigl(r_1^2+r_2^2\bigr) + r_{12}^{-1}  
\end{equation}
is given by\cite{3,5}
\begin{align}\label{a27}
\Psi^{\bullet}(\omega;\vec r_1,\vec r_2) 
  &= \Bigl(\frac{\omega}{\pi}\Bigr)^{3/2}
    \exp\Bigl(-\sfrac{1}{2}\omega \bigl(r_1^2+r_2^2\bigr)\Bigr) \nonumber \\
  &\pheq \times\Bigl[1+ (2\omega)^{-1/2} \;
    \mathfrak{F}\Bigl(\omega/2)^{1/2}r_{12}\Bigr)\nonumber\\
  &\pheq\hphantom{\times\Bigl[}\mbox{}
    +\mathcal{O}\bigl(\omega^{-1}\bigr)\Bigr]
\end{align}
at the limit of $\omega \to \infty$.
The function
\begin{align}\label{a28}
  \mathfrak{F}(t)
  &=-2\pi^{-1/2}(1+\ln2)+t^{-1}\bigl[ 1 -\exp\bigl(t^2\bigr)\erfc (t)\bigr]
\nonumber \\
&\pheq\mbox{} +2\int_0^t \exp\bigl(s^2\bigr)\erfc(s) \, ds  
\end{align}
that enters Eq.~(\ref{a27}) has the small-$t$ expansion
\begin{equation}\label{a29}
  \mathfrak{F}(t)
  =-2\pi^{-1/2} \ln2+t - \frac{2}{3}\pi^{-1/2}t^2
  +\frac{1}{6}t^3 + \mathcal{O}\bigl(t^4\bigr) \,,
\end{equation}
which yields
\begin{equation}\label{a30}
\Psi^{\bullet}_{\infty}(\omega;\vec r_1,\vec r_2) 
= \Bigl(\frac{\omega}{\pi}\Bigr)^{3/2}
\exp\Bigl(-\sfrac{1}{2}\omega \bigl(r_1^2+r_2^2\bigr)\Bigr)
\Bigl( 1+ \sfrac{1}{2}r_{12 }\Bigr)
\end{equation}
as an analog of $\Psi_{\infty}(\omega;\vec r_1,\vec r_2)$.

By virtue of the aforementioned general formalism,\cite{21}
it follows from Eq.~(\ref{a30}) that the asymptotic estimates analogous to
$\widetilde\lambda_{nl}$, 
$\widetilde\lambda_{\mathfrak{n}}$, and $\widetilde v_{nl}$ read\cite{24}
\begin{align}\label{a31}
  \widetilde\lambda^{\bullet}_{nl}
  &= - \frac{4}{\pi^{5/2}}\,\omega^{-1/2}n^{-4}  \,,
\\\label{a32}
  \widetilde\lambda^{\bullet}_{\mathfrak{n}}
  &= - \frac{2^{14/3}}{3^{10/3}\pi^{7/6}}\,\omega^{-1/2} \mathfrak{n}^{-4/3}  \,,
\\\label{a33}
  \tilde t^{\bullet}_{nl}
  &= \frac{\pi/2}{3^{1/2}}\,\omega n^2  \,,
\intertext{and}
\label{a34}
\tilde v^{\bullet}_{nl} &= \omega \,,
\end{align}
respectively.
The large-$n$ asymptotic estimate $\widetilde u^{\bullet}_{nl}$ of the
$m$-independent expectation value  
$u^{\bullet}_{nl} = \langle \psi^{\bullet}_{nlm}(\omega;\vec r) |$
$\exp\bigl(-\frac{1}{2}\omega r^2\bigr)|
\psi^{\bullet}_{nlm}(\omega;\vec r) \rangle$ that enters the analog
\begin{equation}\label{a35}
  \widetilde\lambda^{\bullet}_{nl}
  = - \frac{\omega^{3/2}} {\pi^{1/2}}
  \biggl( \frac{\tilde u^{\bullet}_{nl}}
{\tilde t^{\bullet}_{nl}-\frac{1}{4}\tilde v^{\bullet}_{nl}} \biggr)^2
\end{equation}
of Eq. (\ref{a12}) equals $3^{-1/2}$, whereas those of
$\{ \psi^{\bullet}_{nlm}(\omega;\vec r) \}$ are given by\cite{21,24}
\begin{align}\label{a36}
  \widetilde \psi^{\bullet}_{nlm}(\omega;\vec r)
  &= \Bigl(\frac{4\omega}{\pi}\Bigr)^{1/4}
    J_{l+\frac{3}{2}}(\chi_{nl})^{-1} 
    \frac{\exp\bigl(-\frac{1}{8}\omega r^2\bigr)}{r}\nonumber \\
  &\pheq\times
    \erf \bigl(\omega^{1/2}r/2\bigr)^{1/2} \nonumber \\
  &\pheq\times J_{l+\frac{1}{2}} \Bigl( \chi_{nl}
    \erf\bigl(\omega^{1/2}r/2\bigr) \Bigr) Y_{lm}(\theta,\varphi) \,.
\end{align}

Comparing Eqs.~(\ref{a19}) and (\ref{a36}) reveals that 
$\widetilde \psi_{nlm}(\omega;\vec r)$
$=  2^{3/4} \widetilde \psi^{\bullet}_{nlm}(\omega;2^{1/2} \vec r)$
for all the $n$, $l$, and $m$.
The identities ${\widetilde t_{nl} = 2\widetilde t^{\bullet}_{nl}}$,
${\widetilde v_{nl} = \frac{1}{2}\widetilde v^{\bullet}_{nl}}$,
and ${u_{nl} =u^{\bullet}_{nl}}$ trivially follow from this relationship.

The collective occupancy (per spin and $m$)
$\eta^{\bullet}_l = \sum_{n=1}^\infty \nu^{\bullet}_{nl}$, where 
${\{ \nu^{\bullet}_{nl} \} \equiv \{(\lambda^{\bullet}_{nl})^2 \}}$
is readily obtained from the second-order perturbation theory,\cite{5,23}
which produces  
${\eta^{\bullet}_0 = 1 -(5-2 \, \pi+2 \, \ln 2)/(2 \, \pi)\, \omega^{-1}}$ and
the expression  
\begin{align}\label{a37}
  \eta^{\bullet}_l
  &= \frac{4^{-l}}{2\pi l^2 (2l+1)^2}  
    \Biggl[(2l+1) \,
    _3F_2\biggl(\begin{array}{c}l,l,l+\frac{1}{2}\\[0.5ex]l+1,2l+2\end{array}
  \bigg| 1\biggr) \nonumber\\   
&\pheq\hphantom{\frac{4^{-l}}{2\pi l^2 (2l+1)^2}\Bigg[}\mbox{}
 - 2l\,_3F_2\biggl(\begin{array}{c}l,l+\frac{1}{2},l+\frac{1}{2}\\[0.5ex]
   l+\frac{3}{2},2l+2\end{array}
  \bigg| 1\biggr)\Biggr] \, \omega^{-1}  
\end{align}
valid for ${l \ne 0}$ that involves the hypergeometric functions.
These occupancies exhibit the large-$l$ asymptotic be\-hav\-ior of 
${\lim_{l \to \infty} \, \bigl(l+\frac{1}{2}\bigr)^{\!7} \, \eta^{\bullet}_l
  =15/(32\pi) \, \omega^{-1}}$.

\section{Numerical analysis}\label{sec:num}
It is instructive to juxtapose the predictions pre\-sent\-ed in section
\ref{sec:th-A} against the results of numerical calculations.
Of particular interest is the data computed for $l=0$ that pertains to the
$s$-type NOs.
Highly ac\-cu\-rate approximations of these NOs are provided by the linear
combinations 
\begin{equation}\label{a38}
  \psi_{n00}(\omega;\vec r) \approx
  \sum_{p=0}^{\mathfrak{N}} D_{n0,p}^{(\mathfrak{N})} \mathfrak{f}_{p00}(\omega;r)
\end{equation}
of the square-normalized, i.e.\
$\int_0^{\infty}\bigl|\mathfrak{f}_{p00}(\omega;r)\bigr|^24\pi r^2\,dr=1$
for all $p$, basis functions  
\begin{align}\label{a39}
  \mathfrak{f}_{p00}(\omega;r)
  &= \Bigl( \frac{\omega}{\pi} \Bigr)^{3/4}
    \biggl[\frac{(2p)!!}{(2p+1)!!} \biggr]^{1/2}
    L_{p}^{1/2} (\omega r^2)\nonumber\\  
  &\pheq\times\exp \Bigl(-\sfrac{1}{2}\omega  r^2 \Bigr)   
\end{align}
that are the eigenfunctions of the core Hamiltonian
${-\frac{1}{2}\grad^2+\frac{1}{2}\omega^2r^2}$ and involve the generalized
Laguerre polynomials.
The linear expansion coefficients
$\bigl\{ D_{n0,p}^{(\mathfrak{N})}\bigr\}$ are the eigenvectors of the matrix
$\mathbf{G^{(\mathfrak{N})}}$ with the elements\cite{24} 
\begin{equation}\label{a40}
  G^{(\mathfrak{N})}_{pq}
  = \Bigl( \frac{2}{\pi} \Bigr)^{1/2} 2^{-(p+q)}
  \frac{(2p+2q-1)!!}{\bigl[ (2p+1)! \, (2q+1)!\bigr]^{1/2}} \, ,
\end{equation}
whereas the natural amplitudes $\{ \lambda_{n0} \}$ are approximated by the
corresponding eigenvalues.
For large $p$, 
$G^{(\mathfrak{N})}_{pp} \sim (2 \, \pi)^{-1} \, p^{-3/2}$.

The computation of the quantities pertaining to the ${\omega \to \infty}$
limit of the two-electron harmonium atom in\-volves analogous steps.
For $n \ne 1$,
\begin{equation}\label{a41}
  \psi^{\bullet}_{n00}(\omega;\vec r) \approx
  \sum_{p=1}^{\mathfrak{N}} D_{n0,p}^{\bullet \, (\mathfrak{N})}
  \mathfrak{f}_{p00}(\omega;r)  \,,
\end{equation}
where the linear expansion coefficients
$\bigl\{ D_{n0,p}^{\bullet \, (\mathfrak{N})}\bigr\}$ are the eigenvectors of
the matrix $\mathbf{G^{\bullet \, (\mathfrak{N})}}$ with the elements
\begin{equation}\label{a42}
  G^{\bullet \, (\mathfrak{N})}_{pq}
  = - 
  \frac{(2/\pi)^{1/2}4^{-(p+q)}\,(2p+2q-1)!}
  {\bigl[(2p+1)! \, (2q+1)!\bigr]^{1/2} \, (p+q)!} \, \omega^{-1/2} 
\end{equation}
and the natural amplitudes $\{ \lambda^{\bullet}_{n0} \}$ are approximated by
the corresponding eigenvalues.\cite{25,26}
For large $p$, 
$G^{\bullet\,(\mathfrak{N})}_{pp}\sim-(8\pi)^{-1}\omega^{-1/2}p^{-5/2}$.

\begin{table}[t]
  \caption{\label{tbl:1}The properties of the twenty $s$-type NOs with the
    largest occupation numbers.\footnote{The approximate values computed for
      $\mathfrak{N}=2000$.}} 
\begin{ruledtabular}\begin{tabular}{rrrrr}
  $n$ & \multicolumn{1}{c}{$\lambda_{n0}$}
      & \multicolumn{1}{c}{$\omega^{-1}\, v_{n0}$} 
      & \multicolumn{1}{c}{$\omega^{-1} \, t_{n0}$}  
      & \multicolumn{1}{c}{$u_{n0}$}\\
\hline \noalign{\smallskip}
 1 & 8.408176 $\cdot 10^{-1}$ & 0.500000 &   5.526343 & 0.564894\\
 3 & 4.951911 $\cdot 10^{-2}$ & 0.500000 &  13.563889 & 0.573311\\
 4 & 2.625255 $\cdot 10^{-2}$ & 0.500000 &  25.239319 & 0.575578\\
 5 & 1.624173 $\cdot 10^{-2}$ & 0.500000 &  40.548811 & 0.576457\\
 6 & 1.103445 $\cdot 10^{-2}$ & 0.500000 &  59.490505 & 0.576870\\
 7 & 7.983664 $\cdot 10^{-3}$ & 0.500000 &  82.063317 & 0.577088\\
 8 & 6.043756 $\cdot 10^{-3}$ & 0.500000 & 108.266547 & 0.577213\\
 9 & 4.734019 $\cdot 10^{-3}$ & 0.500000 & 138.099709 & 0.577288\\
10 & 3.808275 $\cdot 10^{-3}$ & 0.500000 & 171.562448 & 0.577335\\
11 & 3.129825 $\cdot 10^{-3}$ & 0.500000 & 208.654495 & 0.577365\\
12 & 2.617812 $\cdot 10^{-3}$ & 0.500000 & 249.375639 & 0.577385\\
13 & 2.221921 $\cdot 10^{-3}$ & 0.500000 & 293.725713 & 0.577398\\
14 & 1.909512 $\cdot 10^{-3}$ & 0.500000 & 341.704579 & 0.577407\\
15 & 1.658657 $\cdot 10^{-3}$ & 0.500000 & 393.312122 & 0.577413\\
16 & 1.454183 $\cdot 10^{-3}$ & 0.500000 & 448.548248 & 0.577416\\
17 & 1.285323 $\cdot 10^{-3}$ & 0.500000 & 507.412875 & 0.577418\\
18 & 1.144259 $\cdot 10^{-3}$ & 0.500000 & 569.905931 & 0.577419\\
19 & 1.025208 $\cdot 10^{-3}$ & 0.500000 & 636.027357 & 0.577419\\
20 & 9.238179 $\cdot 10^{-4}$ & 0.500000 & 705.777098 & 0.577419
\end{tabular}\end{ruledtabular}
\end{table}

Although the basis functions (\ref{a38}) give rise to readily evaluable matrix
elements (\ref{a39}) and (\ref{a41}), the convergence of the computed data
with the basis set size $\mathfrak{N}$ is slow due to the smallness of the
exponents ($\frac{3}{2}$ and $\frac{5}{2}$, respectively) in the
aforementioned power laws governing the decays of $G^{(\mathfrak{N})}_{pp}$
and $G^{\bullet \, (\mathfrak{N})}_{pp}$ with $p$.
Consequently, expansions involving several thousands of these basis functions
are required to produce reasonably accurate properties for tens of the NOs
with the largest occupation numbers.
A proper control of the concomitant roundoff errors requires the employment of
arbitrary-precision arithmetic software.\cite{27} 

\begin{figure}
  \centering
  \includegraphics[width=0.9\columnwidth]{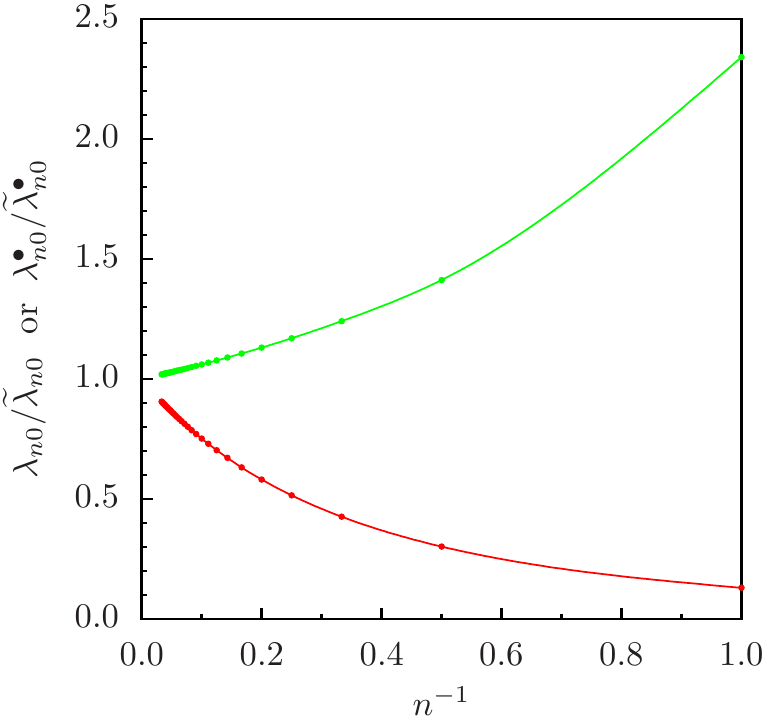}
  \caption{\label{fig:1}%
    The $\lambda_{n0}/\widetilde\lambda_{n0}$ (green) and
    $\lambda_{n0}^{\bullet}/\widetilde\lambda_{n0}^{\bullet}$ (red)
    ratios vs.\ $n^{-1}$ for $ 1\le n \le 30$.
    The lines are provided for eye guidance only.}
\end{figure}

\begin{figure}
  \centering
  \includegraphics[width=0.9\columnwidth]{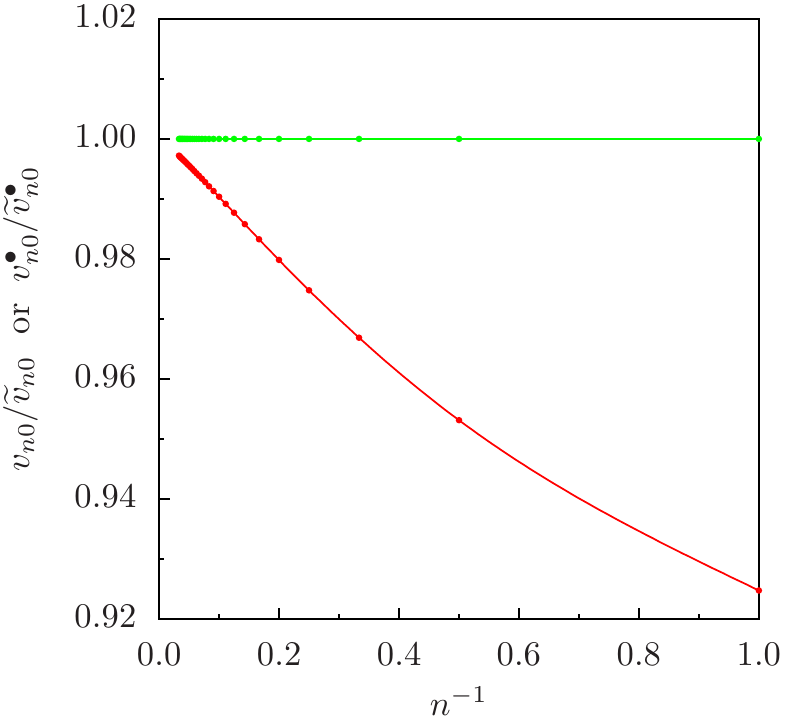}
  \caption{\label{fig:2}%
    The $v_{n0}/\widetilde v_{n0}$ (green) and
    $v_{n0}^{\bullet}/\widetilde v_{n0}^{\bullet}$ (red) ratios vs.\
    $n^{-1}$ for $ 1\le n \le 30$.
    The lines are provided for eye guidance only.}
\end{figure}

\begin{figure}[t]
  \centering
  \includegraphics[width=0.9\columnwidth]{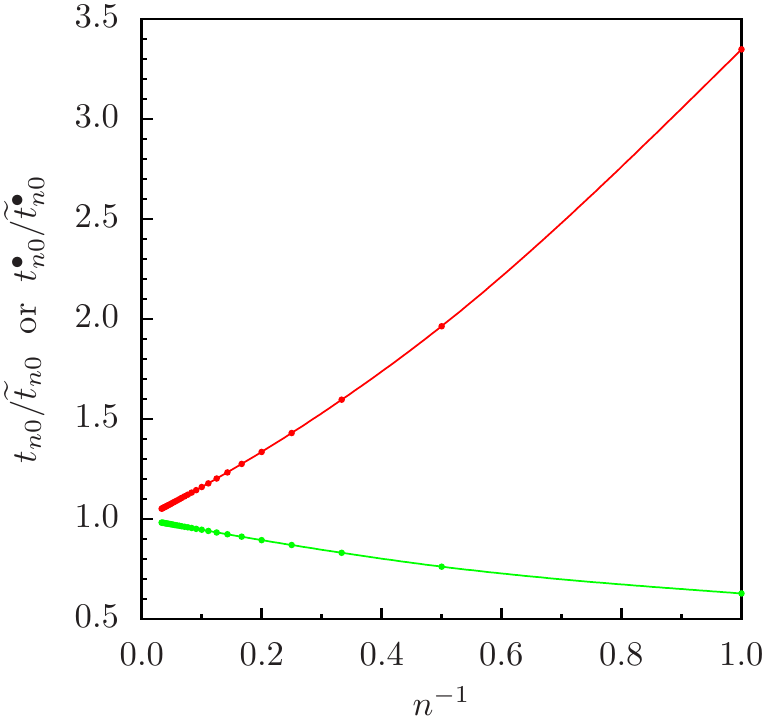}
  \caption{\label{fig:3}%
    The $t_{n0}/\widetilde t_{n0}$ (green) and
    $t_{n0}^{\bullet}/\widetilde t_{n0}^{\bullet}$ (red) ratios vs.\ $n^{-1}$
    for $ 1\le n \le 30$.
    The lines are provided for eye guidance only.}
\end{figure}

\begin{figure}[t]
  \centering
  \includegraphics[width=0.9\columnwidth]{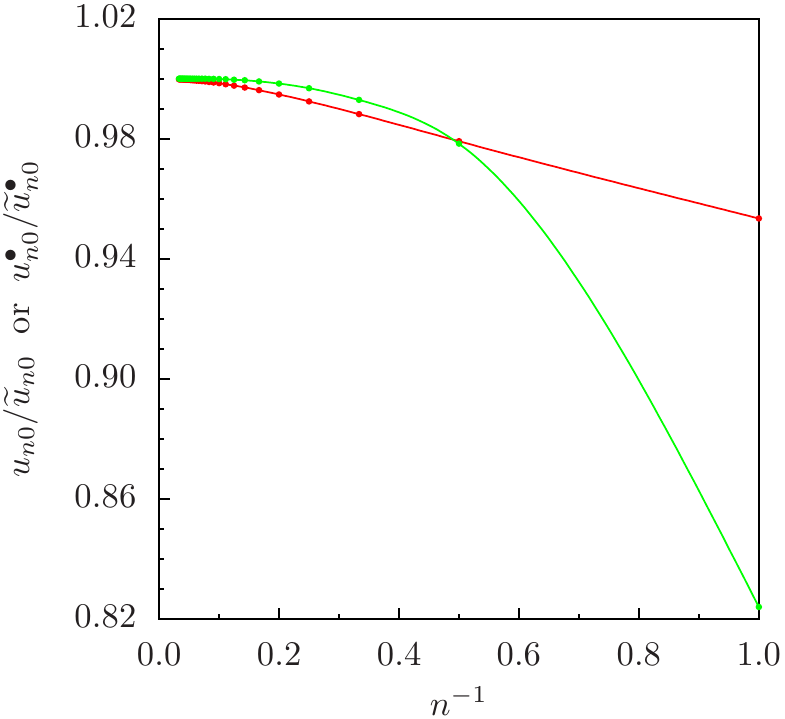}
  \caption{\label{fig:4}%
    The $u_{n0}/\widetilde u_{n0}$ (green) and
    $u_{n0}^{\bullet}/\widetilde u_{n0}^{\bullet}$ (red) ratios vs.\ $n^{-1}$
    for $ 1\le n \le 30$.
    The lines are provided for eye guidance only.}
\end{figure}

In Table \ref{tbl:1}, the highly accurate values of $\{ \lambda_{n0} \}$,
$\{\omega^{-1}$ $v_{n0}\}$, $\{\omega^{-1}t_{n0}\}$, and $\{u_{n0}\}$ for 
${1 \le n \le 20}$ are com\-piled, whereas the corresponding
$\{{\lambda_{n0}}/{\widetilde\lambda_{n0}}\}$, 
$\{{v_{n0}}/{\widetilde v_{n0}}\}$,
$\{{t_{n0}}/{\widetilde t_{n0}}\}$, and
$\{{u_{n0}}/{\widetilde u_{n0}}\}$ ratios are displayed in
Figs.~\ref{fig:1}--\ref{fig:4} together with their counterparts for the
two-electron harmonium atom at the $\omega \to \infty$ limit.
Inspection of \mbox{these} figures confirms the constancy of $v_{n0}$ with
respect to $n$ and the rapid rates at which  
$\lambda_{n0}$, $t_{n0}$, $u_{n0}$, $\lambda^{\bullet}_{n0}$,
$v^{\bullet}_{n0}$, $t^{\bullet}_{n0}$, and $u^{\bullet}_{n0}$ approach their
respective large-$n$ asymptotics.

\begin{figure*}[t]
  \centering
  \begin{tabular}{@{}c@{\qquad}c@{}}
  a)\hspace*{-1.5em} \raisebox{-\height}{\includegraphics[width=0.9\columnwidth]%
      {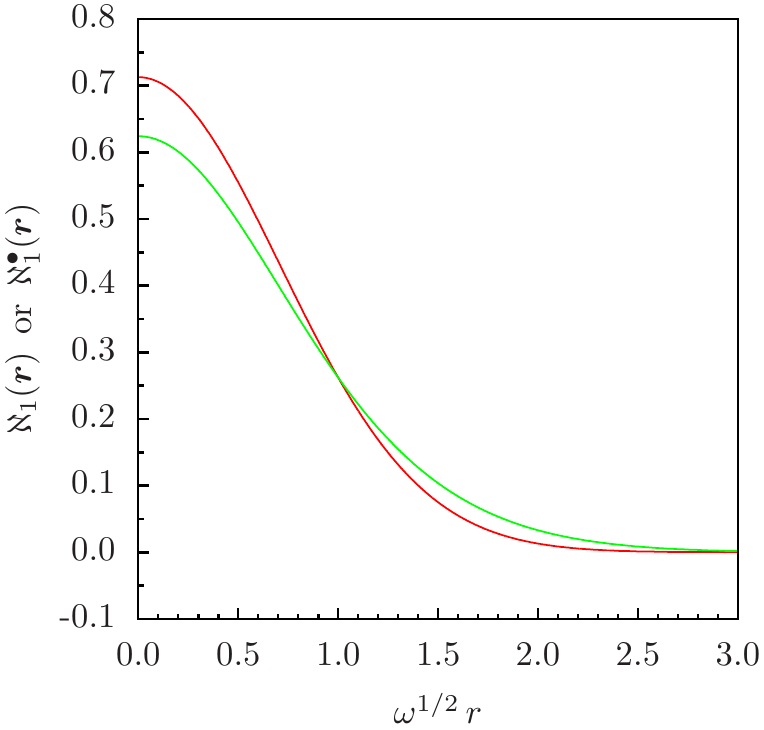}}&
  b)\hspace*{-1.5em} \raisebox{-\height}{\includegraphics[width=0.9\columnwidth]%
      {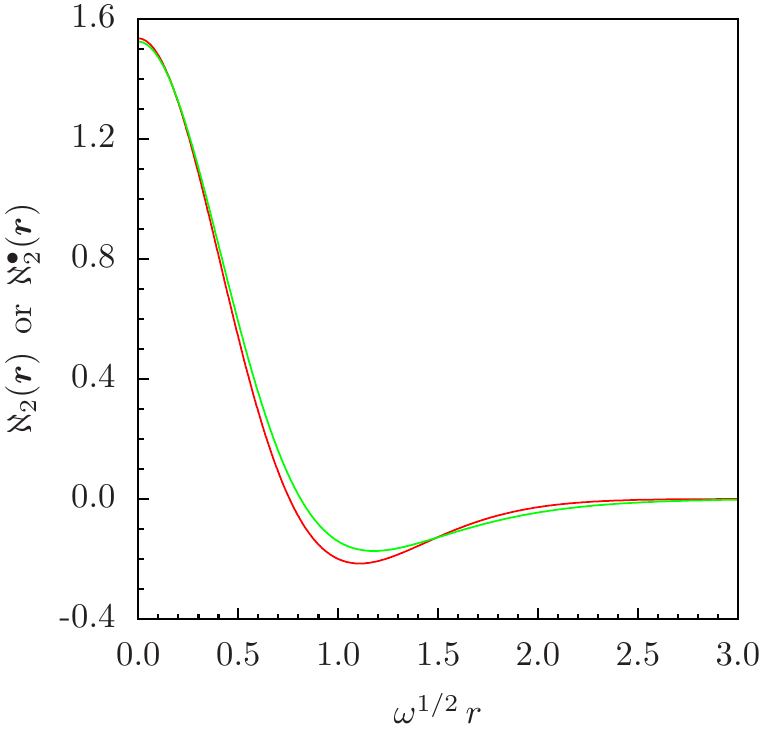}}\\
  c)\hspace*{-1.5em} \raisebox{-\height}{\includegraphics[width=0.9\columnwidth]%
      {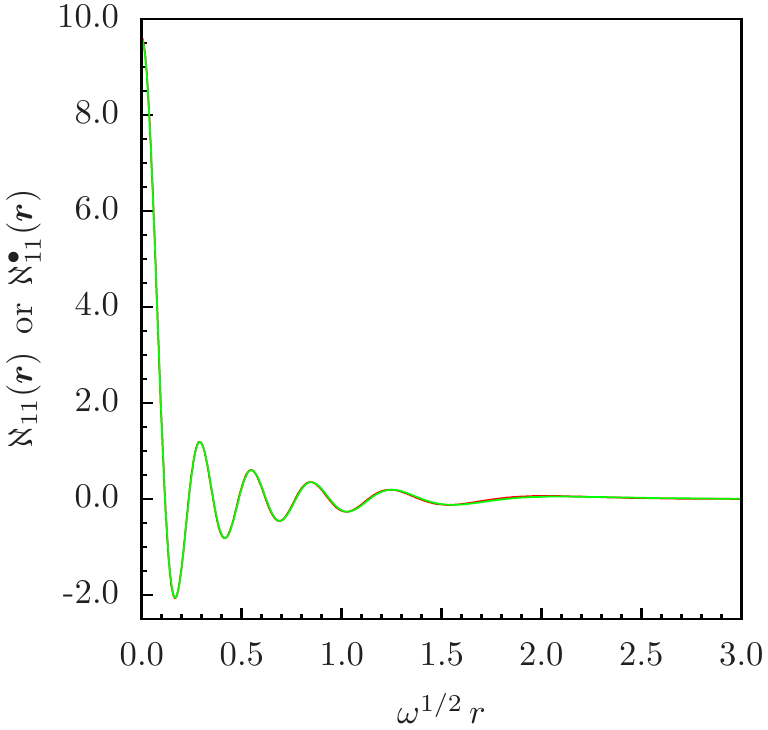}}&
  d)\hspace*{-1.5em} \raisebox{-\height}{\includegraphics[width=0.9\columnwidth]%
      {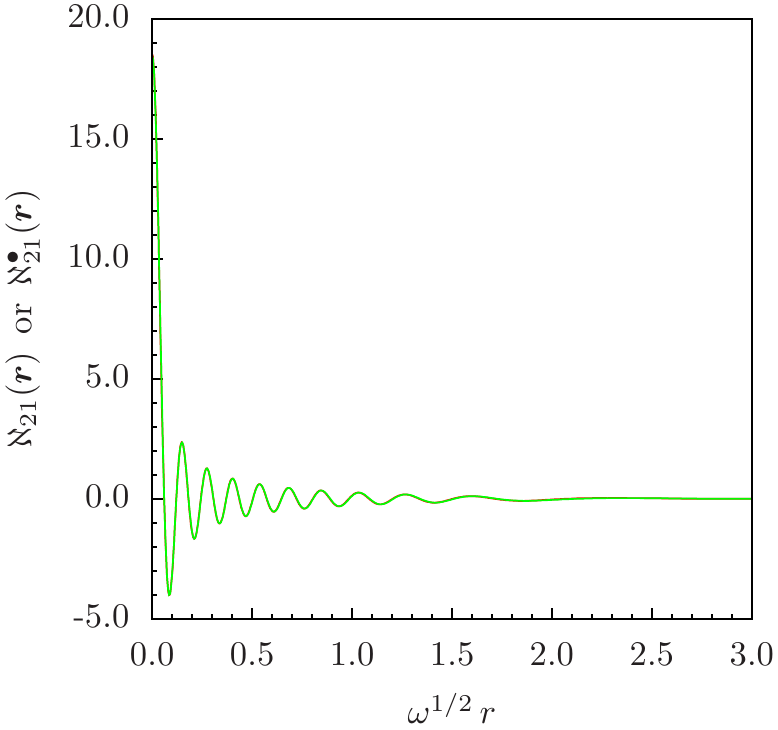}}
  \end{tabular}
  \caption{\label{fig:5}%
    The scaled NOs
    $\aleph_n(r)\equiv\omega^{-3/4}\psi_{n00}(\omega;\vec r)$ (green) and 
    $\aleph^{\bullet}_n(r)\equiv(\omega/2)^{-3/4}%
    \psi^{\bullet}_{n00}(\omega;2^{1/2}\vec r)$ (red) vs.\
    $\omega^{1/2}r$ for a) $n=1$, b) $n=2$, c) $n=11$, and d) $n=21$.}
\end{figure*}

The plots of the scaled NOs presented for $n=1$ [\mbox{note} that
${\psi^{\bullet}_{100}(\omega;\vec r)=\mathfrak{f}_{000}(\omega;r)}$] and ${n=2}$
in Figs.~\ref{fig:5}a and \ref{fig:5}b are quite similar.
At larger $n$ (Figs.~\ref{fig:5}c and \ref{fig:5}d), they become virtually
indistinguishable, demonstrating the remark\-able accuracy of the approximate
identity $\psi_{n00}(\omega;\vec r) \approx
2^{3/4}\psi^{\bullet}_{n00}(\omega;2^{1/2}\vec r)$ that follows from the
asymptotic estimates (\ref{a19}) and (\ref{a36}).
Although, strictly speaking, these estimates are valid only at the $n \to
\infty$ limit, the approximate identity appears to be closely followed already
for $n=11$.

\section{Discussion and conclusions}
There are many peculiarities inherent in the one-particle description of the
contactium.
Some of them follow directly from the infinite values of the kinetic and
interparticle interaction energies.
Thus, combining Eqs.~(\ref{a11}), (\ref{a12}), (\ref{a14}), and (\ref{a18})
yields the asymptotic power law 
\begin{equation}\label{a43}
\lim_{\mathfrak{n} \to \infty}\mathfrak{n}^{2/3}\nu_{\mathfrak{n}}t_{\mathfrak{n}}
= \lim_{\mathfrak{n} \to \infty}\frac{2\omega}{(3\pi)^{1/2}}\mathfrak{n}^{2/3}
 \lambda_{\mathfrak{n}} = \frac{2^{10/3}}{3^{13/6}\pi^{4/3}}\,\omega 
\end{equation}
for the contribution $\nu_{\mathfrak{n}}t_{\mathfrak{n}}$ of the
$\mathfrak{n}$th NO to the kinetic energy.
Consequently, the sum 
$\sum_{\mathfrak{n}=1}^{\infty}\nu_{\mathfrak{n}}t_{\mathfrak{n}}$ diverges,
as expected.
On the other hand, the kinetic energy $T_{\rm{KS}}$ of the fictitious
noninteracting system involved in the description of the contactium within the
Kohn--Sham formalism\cite{28} is finite, i.e.\ 
\begin{equation}\label{a44}
  T_{\rm{KS}}
  = \frac{1}{2}\int \bigl|\grad\sqrt{\rho_{\infty}(\omega;\vec r)}\bigr|^2 \,
  d^3 \vec r  \approx 1.130 \, 576 \; \omega 
\end{equation}
where [compare with Eq.~(\ref{a22})]
\begin{equation}\label{a45}
  \rho_{\infty}(\omega;\vec r)
  =  \frac{1}{\pi} \omega  \exp\bigl(-2\omega r^2\bigr)
  \frac{\erfi\bigl(\omega^{1/2} \, r\bigr)}{r}
\end{equation}
is the one-particle density (per spin).

In analogy to that of a Coulombic system, the interparticle interaction energy
of the contactium can be formally partitioned into the direct, exchange, and
correlation contributions.
For the first two of those, one obtains
\begin{align}\label{a46}
  J&=2\int\hspace{-4pt}\int\rho_{\infty}(\omega;\vec r_1)
     \,\rho_{\infty}(\omega;\vec r_2) \,\delta_{\rm{reg}}(\vec r_{12})
     \, d^3 \vec r_1 \, d^3 \vec r_2 \nonumber \\ 
  &= 2\int\rho_{\infty}(\omega;\vec r)^2 \, d^3 \vec r 
   = \frac{4}{\pi^{3/2}}\arctan_var(2^{-3/2}) \; \omega^{3/2} 
\end{align}
and
\begin{align}\label{a47}
  K &= - \int\hspace{-4pt}\int
      \bigl|\Gamma_{\infty}(\omega;\vec r_{1},\vec r_2)\bigr|^2 \, 
\delta_{\rm{reg}}(\vec r_{12}) \, d^3 \vec r_1 \, d^3 \vec r_2\nonumber \\
&= - \int\rho_{\infty}(\omega;\vec r)^2 \, d^3 \vec r  
=- \frac{2}{\pi^{3/2}} \arctan_var(2^{-3/2}) \, \omega^{3/2} \,,
\end{align}
respectively.
Although these contributions are finite-valued, they scale like $\omega^{3/2}$
rather than $\omega$, as would be expected from the overall scaling of the
energy.  
The correlation contribution is both negative and infinite.

For Coulombic systems, the 1-RDMFT formalism in\-volves only a single
component (i.e.\ the correlation part of the electron-electron repulsion
energy) of the total energy that is given by an unknown functional.
In contrast, there are three energy components (i.e.\ the kinetic energy
together with the exchange and correlation part of the electron-electron
repulsion energy) within DFT for which one has to resort to approximate
expressions.
In the case of the contactium, this advantage  enjoyed by 1-RDMFT over DFT is
lost as the kinetic and interparticle inter\-action contributions to the total
energy can no longer be considered separately.
On the other hand, taking into account the aforementioned finite-valuedness of
$T_{\rm{KS}}$, one expects a reasonable description of many-particle \mbox{analogs}
of the contactium with the Kohn--Sham approach, pro\-vid\-ed a suitable functional
is constructed. 

The present study leads to the somewhat surprising conclusion that almost
identical natural orbitals can pertain to two systems with diametrically
different interparticle interactions giving rise to entirely different
behav\-ior of the respective wave functions at the spatial two-particle
coalescences.
It thus appears that the gross of the information about these interactions is
contained in the occupation numbers or, to be more precise, in \mbox{their}
asymptotic dependence on the ordinal number.
This observation strongly suggests that quantitative measures of particle
correlation based upon one-particle quantities should be constructed from the
occupation numbers rather than properties of the corresponding natural
or\-bitals. 

The unusual properties of contactium are bound to stimulate further research
on strongly correlated systems.
Of particular interest is the extension of the present  study to species
involving large numbers of either fermions or bosons subject to various
confining potentials.

\section*{Acknowledgments}
The research described in this publication has been funded by the National
Science Center (Poland) under grant \mbox{2018/31/B/ST4/00295} and supported
by the National Research Foundation, Singapore and A*STAR under its CQT
Bridging Grant and its Quantum Engineering Programme
(grant \mbox{NRF2022-QEP2-02-P16} supports J.H.H.).
One of the authors (J.C.) thanks the good people of CQT for their splendid
hospitality during his stay in Singapore.


\begin{thebibliography}{99}
\bibitem{1} T. Helgaker, P. J{\o}rgensen, J. Olsen,
  \textit{Molecular Electronic-Structure Theory}
  (John Wiley \& Sons, Chichester, 2000).
\bibitem{2} M. Taut, \textit{Phys. Rev. A} \textbf{48}, 3561 (1993);
  \textit{J. Phys. A} \textbf{27}, 1045 (1994);
  \textbf{27}, 4723(E) (1994).
\bibitem{3} J. Cioslowski and K. Pernal,
  \textit{J. Chem. Phys.} \textbf{113}, 8434 (2000).
\bibitem{4} H. F. King, \textit{Theor. Chim. Acta} \textbf{94}, 345 (1996). 
\bibitem{5} R. J. White and W. Byers Brown,
  \textit{J. Chem. Phys.} \textbf{53}, 3869 (1970).
\bibitem{6} N. R. Kestner and O. Sinano\={g}lu,
  \textit{Phys. Rev.} \textbf{128}, 2687 (1962).
  \newline E. Santos, \textit{Anal. R. Soc. Esp. Fis. Quim.}
  \textbf{64}, 177 (1968).
\bibitem{7} C. Filippi, C. J. Umrigar, and M. Taut,
  \textit{J. Chem. Phys.} \textbf{100}, 1290 (1994).
  \newline S. Ivanov, K. Burke, and M. Levy,
  \textit{ibid.} \textbf{110}, 10262 (1999).
  \newline  M. Taut, A. Ernst, and H. Eschrig,
  \textit{J. Phys. B} \textbf{31}, 2689 (1998).
  \newline P. Gori-Giorgi and A. Savin,
  \textit{Int. J. Quantum Chem.} \textbf{109}, 2410 (2009).
\bibitem{8} M. Rodr\'{\i}guez-Mayorga, E. Ramos-Cordoba, M. Via-Nadal,
  M. Piris, and E. Matito,
  \textit{Phys. Chem. Chem. Phys.} \textbf{19}, 24029 (2017). 
  \newline K. J. H. Giesbertz and R. van Leeuwen,
  \textit{J. Chem. Phys.} \textbf{139}, 104110 (2013).
  \newline S. Crisostomo, M. Levy, and K. Burke,
  \textit{J. Chem. Phys.} \textbf{157}, 154106 (2022).
  \newline D. P. Kooi and P. Gori-Giorgi,
  \textit{Theor. Chem. Acc.} \textbf{137}, 166 (2018).
  \newline  S. \'{S}miga, F. D. Sala, P. Gori-Giorgi, and E. Fabiano,
  \textit{J. Chem. Theory Comput.} \textbf{18}, 5936 (2022).
\bibitem{9} P.-F. Loos and P. M. W. Gill, \textit{Phys. Rev. Lett.}
  \textbf{103}, 123008 (2009).
\bibitem{10} P.-F. Loos and P. M. W. Gill,
  \textit{J. Chem. Phys.} \textbf{132}, 234111 (2010).
  \newline J. Jung and J. E. Alvarellos,
  \textit{J. Chem. Phys.} \textbf{118}, 10825 (2003).
  \newline D.C. Thompson and A. Alavi,
  \textit{Phys. Rev. B} \textbf{66}, 235118 (2002);
  \textbf{68}, 039901(E) (2003).
\bibitem{11} For a recent review see: K. Pernal and K. J. H. Giesbertz,
  \textit{Top. Curr. Chem.} \textbf{368}, 125 (2016).
\bibitem{12} C. L. Benavides-Riveros, J. Wolff, M. A. L. Marques, and
  C. Schilling,
  \textit{Phys. Rev. Lett.} \textbf{124}, 180603 (2020).
\bibitem{13} J. Cioslowski, \textit{J. Chem. Theory Comput.}
  \textbf{16}, 1578 (2020).
\bibitem{14} E. Fermi, \textit{Ricerca Scient.} \textbf{7}, 13 (1936).
\bibitem{15} K. Huang and C. N. Yang,
  \textit{Phys. Rev.} \textbf{105}, 767 (1957).
\bibitem{16} For a review, see for example: D. Blume,
  \textit{Rep. Prog. Phys.} \textbf{75}, 046401 (2012).
\bibitem{17} T. Busch, B.-G. Englert, K. Rz\k{a}\.zewski, and M. Wilkens,
  \textit{Found. Phys.} \textbf{28}, 549 (1998).
\bibitem{18} P.-O. L\"owdin and H. Shull, \textit{Phys. Rev.}
  \textbf{101}, 1730 (1956).
\bibitem{19} K. Chadan, \textit{Il Nuovo Cimento A}, \textbf{58}, 191 (1968).
\bibitem{20} A. Martin, \textit{Helv. Phys. Acta} \textbf{45}, 140 (1972);
  H. Tamura, \textit{Proc. Japan Acad.} \textbf{50}, 19 (1974). 
\bibitem{21} J. Cioslowski and K. Strasburger,
  \textit{J. Chem. Theory Comput.} \textbf{17}, 6918 (2021).
\bibitem{22} J. Cioslowski and F. Pr\k{a}tnicki,
  \textit{J. Chem. Phys.} \textbf{150}, 074111 (2019).
\bibitem{23} J. Cioslowski, \textit{Theor. Chem. Acc.} \textbf{134}, 113 (2015).
\bibitem{24} J. Cioslowski, \textit{J. Chem. Phys.} \textbf{148}, 134120 (2018).
\bibitem{25} J. Cioslowski, \textit{Theor. Chem. Acc.} \textbf{137}, 173 (2018).
\bibitem{26} J. Cioslowski and F. Pr\k{a}tnicki,
  \textit{J. Chem. Phys.} \textbf{151}, 184107 (2019).
\bibitem{27} Mathematica, Version 12.2.0.0,
  Wolfram Research, Inc., Champaign, IL, 2020.
\bibitem{28} W. Kohn,  and L. J. Sham, \textit{Phys. Rev.}, A1133 (1965).
\end{thebibliography}
\end{document}